\documentclass[lettersize,journal]{IEEEtran}
\usepackage{amsmath,amsfonts,amssymb}
\usepackage{graphicx}
\usepackage{algorithmicx}
\usepackage{algorithm}
\usepackage{algpseudocode}
\usepackage{textcomp}
\usepackage{makecell}
\usepackage{multirow,multicol}
\usepackage[colorlinks]{hyperref}
\hypersetup{citecolor=blue}
\usepackage[table]{xcolor}
\usepackage{threeparttable}
\usepackage{balance}
\usepackage{booktabs}
\usepackage[mathscr]{eucal}
\begin{document}

\title{A Universal Block Error Rate Bound\\for Fluid Antenna Systems}
\author{Zhentian Zhang, David Morales-Jimenez, Hao Jiang,  Christos Masouros
\thanks{}
\thanks{Z. Zhang is with the National Mobile Communications Research Laboratory, Southeast University, Nanjing, 210096, China. (e-mail: zhentianzhangzzt@gmail.com).}
\thanks{D. Morales-Jimenez is with the Department of Signal Theory, Networking and Communications, University of Granada, Granada 18071, Spain (e-mail: dmorales@ugr.es).}
\thanks{H. Jiang is with the National Mobile Communications Research Laboratory, Southeast University, Nanjing 210096, China , and also the School of Artificial Intelligence, Nanjing University of Information Science and Technology, Nanjing 210044, China (e-mail: jianghao@nuist.edu.cn).}
\thanks{C. Masouros is with the Department of Electronic and Electrical Engineering, University College London, Torrington Place, WC1E 7JE, United Kingdom  (e-mails: c.masouros@ucl.ac.uk). }
\thanks{The work of D. Morales-Jimenez is supported in part by the State Research Agency (AEI) of Spain and the European Social Fund under grant RYC2020-030536-I and by MICIU/AEI/10.13039/501100011033 under grant PID2023-149975OB-I00 (COSTUME). The work of H. Jiang is supported in part by the National Natural Science Foundation of China (NSFC) projects (No. 62471238).}}

%

\maketitle
\thispagestyle{empty}
\begin{abstract}
Fluid antenna systems (FASs) offer genuine simplicity for communication network design by eliminating expensive hardware overhead and reducing the complexity of access protocol architectures. Through the discovery of significant spatial diversity within a compact antenna space, FASs enable the implementation of reconfigurable-antenna-based architectures. However, current state-of-the-art studies rarely investigate the impact of finite blocklength constraints on FAS-based designs, leaving a gap in both analytical modeling and the establishment of a solid, universally applicable performance metric for finite blocklength fluid antenna systems (FBL-FAS). In this work, we focus on the study of FBL-FAS and, more importantly, derive a block error rate (BLER) bound that serves as a general and practical performance benchmark across various FAS architectures. The proposed BLER bound is computable both with and without an explicit statistical model, meaning that the BLER performance can be characterized analytically or empirically under model-aware or model-free system scenarios. Moreover, when the statistical model is known, the analytical results derived from the proposed BLER bound exhibit strong alignment with the empirical findings, demonstrating the remarkable simplicity, accuracy, and universality of the proposed BLER bound.
\end{abstract}


\section{Introduction}
Fluid antenna systems (FASs) refer to any software-controllable fluidic, conductive, or dielectric structure capable of dynamically altering its radiation characteristics according to system requirements~\cite{fas_tutorial}. Unlike the current evolutionary trend in wireless systems, from multi-input multi-output (MIMO) and massive MIMO toward future MIMO architectures requiring large swarms of costly radio frequency (RF) units. FASs introduce a fundamentally simpler design philosophy. By exploiting substantial spatial diversity gains within a compact antenna space~\cite{fas-twc-21}, FASs demonstrate that significant spectrum and energy efficiency improvements can be achieved through reconfigurable antenna structures \cite{FAS1,FAS3,kit_electronic}. The concept of reconfigurable fluid antennas has already begun reshaping contemporary communication network designs, influencing areas such as multiple access~\cite{chernoff2,FAS2}, joint communication and sensing~\cite{chernoff3, ISAC_2}, signal processing \cite{SP_FAS} and beyond \cite{JH_FAS}.

However, the finite blocklength effect, a crucial constraint in practical communication system design, has not yet been investigated in the context of potential FAS implementations. Moreover, there is a lack of a solid and universal performance evaluation metric for finite blocklength fluid antenna systems (FBL-FAS). Consequently, this work will introduce the concept of FBL-FAS and investigates the performance of a single-antenna FAS under finite blocklength constraints, focusing on the block error rate (BLER). Fundamentally, a universally applicable BLER bound for FASs is proposed in the spirit of the Chernoff inequality. 

The proposed bound can be computed both analytically (through statistical characterization) and empirically, thereby accommodating a wide range of fluid-antenna-based system designs, including both model-free and model-based FAS architectures. Furthermore, the analytical predictions exhibit strong agreement with empirical results, demonstrating that the proposed BLER bound can be feasibly employed to conserve computational resources that would otherwise be consumed by exhaustive analyses using complex models. By utilizing the proposed BLER bound, the performance of FASs under various channel correlation models can be conveniently evaluated, as system behavior can be explored either via Monte Carlo simulations or through corresponding statistical distribution models. 

{\em The reproducible simulation code for this work can be accessed at https://github.com/BrooklynSEUPHD/Universal-Block-Error-Rate-Bound-for-Fluid-Antenna-Systems.git.}

The remainder of this paper is organized as follows. Section~\ref{sec:system model} introduces the channel model and signal model of FBL-FAS. Section~\ref{Main Results} presents the main theoretical results, while all proofs are provided in Appendix~\ref{appendx}. Section~\ref{sec: numerical results} illustrates the numerical results and evaluates the BLER performance of FBL-FAS, demonstrating that the superiority of FAS is preserved under FBL assumption. Finally, Section~\ref{conclusion} concludes the paper. \textit{Notations:} In this paper, unbolded lowercase letters (e.g., $a$) and uppercase letters (e.g., $A$) denote scalars, while bold lowercase (e.g., $\boldsymbol{a}$) and bold uppercase letters (e.g., $\boldsymbol{A}$) denote vectors and matrices, respectively. The real and complex domains are denoted by $\mathbb{R}$ and $\mathbb{C}$, respectively. Sets are represented by calligraphic letters (e.g., $\mathcal{A}$), with $\mathcal{A}\setminus\mathcal{A}^{\prime}$ denoting the set difference. Commonly used symbols include the expectation $\mathrm{E}[\cdot]$, the $l_2$-norm $\|\cdot\|_2^2$, the conjugate $(\cdot)^{*}$, the transpose $(\cdot)^{\mathrm{T}}$, and the Hermitian transpose $(\cdot)^{\mathrm{H}}$.
\section{System Model}\label{sec:system model}
In this work, we consider a {\em single-antenna} (receiving) base station that serves multiple single-antenna user equipments in an uplink transmission. The users' antenna placement can be dynamically switched among $N$ predefined locations, which are evenly distributed along a linear dimension of length $W\lambda$, where $\lambda$ denotes the wavelength at the given frequency. Each antenna at a given location is referred to as a {\em port} and is modeled as an ideal point antenna. Assuming $U$ active users transmit $M$-length signals and omitting asynchronous issues, the received signal is written as:
\begin{equation}\label{eq:5}
	\boldsymbol{y}= g_{i,k}\boldsymbol{x}_i+\sum_{u\neq i}^{U}g_{u,k}\boldsymbol{x}_u+\boldsymbol{\eta},
\end{equation}
where $\boldsymbol{\eta}$ is the additive white Gaussian noise (AWGN) with zero mean and variance $\frac{\sigma^2_{\eta}}{M},~\|\boldsymbol{\eta}\|^2_2=\sigma^2_{\eta}$ and the signals carrying information are generalized as Gaussian codewords with zero mean and variance $\frac{1}{M}$, i.e., $\{\boldsymbol{x}_u\in \mathbb{C}^{M},~\left \| \boldsymbol{x}_u \right \|_2^2=1\}$. Our goal is to investigate how the limited blocklength $M$ and the user density $U$ affect the performance of the single antenna FAS and the conventional $L$-antenna system in terms of BLER. Moreover, the signal-to-noise ratio (SNR) is defined by:
\begin{equation}
	\mathrm{SNR} = \frac{\mathrm{E}\left[\|g_{i,k}\boldsymbol{x}_i\|_2^2\right]}{\mathrm{E}\left[\|\boldsymbol{\eta}\|_2^2\right]}=\frac{\sigma^2}{\sigma^2_{\eta}}.
\end{equation}
\subsection{Simple Reference Channel Channel}\label{sec.simple_reference_model}
The first port is taken as the reference location, with the displacement of the $k$-th port measured relative to it as:

\begin{equation}\label{eq:1}
	\Delta d_{k,1} = \left(\frac{k-1}{N-1}\right)W\lambda, k=1,2,\ldots,N.
\end{equation}

Let $g_{u,k}$ denote the channel coefficient of the $u$-th user at the $k$-th port following a circularly symmetric complex Gaussian distribution with zero mean and variance $\sigma^2$. Under this model, the channel response amplitude $\rvert g_{u,k} \lvert$ is Rayleigh distributed with PDF:
\begin{equation}\label{eq:2}
	p_{\rvert g_{u,k} \lvert} \left(r\right) = \frac{2r}{\sigma^2}e^{-\frac{r^2}{\sigma^2}},~\text{for}~r\ge 0~\text{with}~\mathrm{E}\left[\rvert g_{u,k} \lvert^2\right]=\sigma^2.
\end{equation}
Furthermore, the channels at the $N$ antenna ports in a FAS are denoted as \cite{fas-twc-21}:
\begin{equation}\label{eq:3}
	\left\{\begin{matrix}
		\begin{aligned}
		g_{u,1}&=\sigma x_{u,0}+j\sigma y_{u,0}\\
		g_{u,k}&=\sigma\left(\sqrt{1-\mu_{k}^{2}}x_{u,k}+\mu_{k}x_{u,0}\right)+\\
		&j\sigma\left(\sqrt{1-\mu_{k}^{2}}y_{u,k}+\mu_{k}y_{u,0}\right)~\mathrm{for}~k=2,\ldots,N,
		\end{aligned}
	\end{matrix}\right.
\end{equation}
where $x_{u,0},x_{u,2},\ldots,x_{u,N},y_{u,0},y_{u,2},\ldots,y_{u,N}$ are all independent Gaussian random variables with zero mean and variance $\frac{1}{2}$, and $\left\{\mu_k\right\}$ are parameters quantifying the ports' correlation with respect to the first port:
\begin{equation}\label{eq:4}
	\mu_k=J_0\left(\frac{2\pi\left(k-1\right)}{N-1}W\right),~\text{for}~k=1,2,\ldots,N,
\end{equation} 
 where $J_0\left(\cdot\right)$ is the zero-order Bessel function of the first kind and $\mu_1=J_0\left(0\right)=1$. The channels $\left\{g_{u,k}\right\}$ are correlated since the ports can be arbitrarily close to each other.
\subsection{Modified Reference Correlation Model}
The correlation model in \eqref{eq:3}, as an early attempt to characterize FAS correlation, has been widely adopted but lacks accuracy because it only considers correlation relative to a reference port and fails to capture the relationship between arbitrary adjacent ports. To address this limitation, a modified reference model was introduced in \cite[Eq.~3]{fas_tutorial}, \cite{kit_electronic}, where the correlation constants in \eqref{eq:3} are replaced with a unified correlation parameter:
\begin{equation}\label{eq:modified}
	\mu_k = \mu = \sqrt{2}\sqrt{\,_1F_2\!\left(\tfrac{1}{2};1,\tfrac{3}{2};-\pi^2W^2\right) - \frac{J_1(2\pi W)}{2\pi W}},
\end{equation}
where $_1F_2(\cdot;\cdot;\cdot)$ is the generalized hypergeometric function and $J_1(\cdot)$ denotes the first-order Bessel function of the first kind. This model connects all ports within the fluid antenna without relying on a reference port, effectively approximating the average squared spatial correlation of a practical fluid antenna array while maintaining analytical tractability.
\subsection{Fully Correlated Channel Model}
A fully correlated channel model provides the most accurate description of inter-port correlation based on channel covariance \cite{fas_tutorial}. Although such modeling guarantees accuracy, it often compromises analytical tractability due to complex statistical dependencies. Nonetheless, the proposed BLER bound remains applicable, as it yields accurate results even with empirically generated samples. Following the construction method in \cite{block_FBL_FAS}, let $\boldsymbol{g}_u \in \mathbb{C}^{N}$ represent the channel coefficient vector of the $u$-th user, and let $\boldsymbol{\Sigma} \in \mathbb{C}^{N \times N}$ denote the corresponding spatial correlation matrix. For uniformly spaced ports, $\boldsymbol{\Sigma}$ follows a Toeplitz structure:
\begin{equation}\label{eq:full}
	\small
	\boldsymbol{\Sigma} =
	\begin{pmatrix}
		a(0) & a(1) & a(2) & \dots & a(N-1)\\
		a(-1) & a(0) & a(1) & \dots & a(N-2)\\
		\vdots & \ddots & \ddots & \vdots & \vdots\\
		a(-N+1) & a(-N+2) & \dots & a(-1) & a(0)
	\end{pmatrix},
\end{equation}
where the generating function is defined as
\begin{equation}
	a(n) = \operatorname{sinc}\!\left(\frac{2\pi n W}{N-1}\right).
\end{equation}
Given $\boldsymbol{\Sigma}$, the channel vector $\boldsymbol{g}_u$ can be efficiently generated using the eigenvalue-based construction~\cite[Eq.~5]{fas_tutorial}:
\begin{equation}\label{eq:channel_eigenvalue}
	\boldsymbol{g}_u = \boldsymbol{Q}\boldsymbol{\Lambda}^{\frac{1}{2}}\boldsymbol{g}_{0},
\end{equation}
where $\boldsymbol{Q}$ is the eigenvector matrix from the decomposition $\boldsymbol{\Sigma} = \boldsymbol{Q}\boldsymbol{\Lambda}\boldsymbol{Q}^{\mathrm{H}}$, and $\boldsymbol{g}_{0} \in \mathbb{C}^{N} \sim \mathcal{CN}\!\left(\boldsymbol{0}, \sigma^2\boldsymbol{I}\right)$.
\section{Main Results}\label{Main Results}
In this section, the universal BLER bound for FAS is established, and detailed derivations are provided in the Appendices.
\begin{figure}[t!]
	\centering
	\includegraphics[width=3.5in]{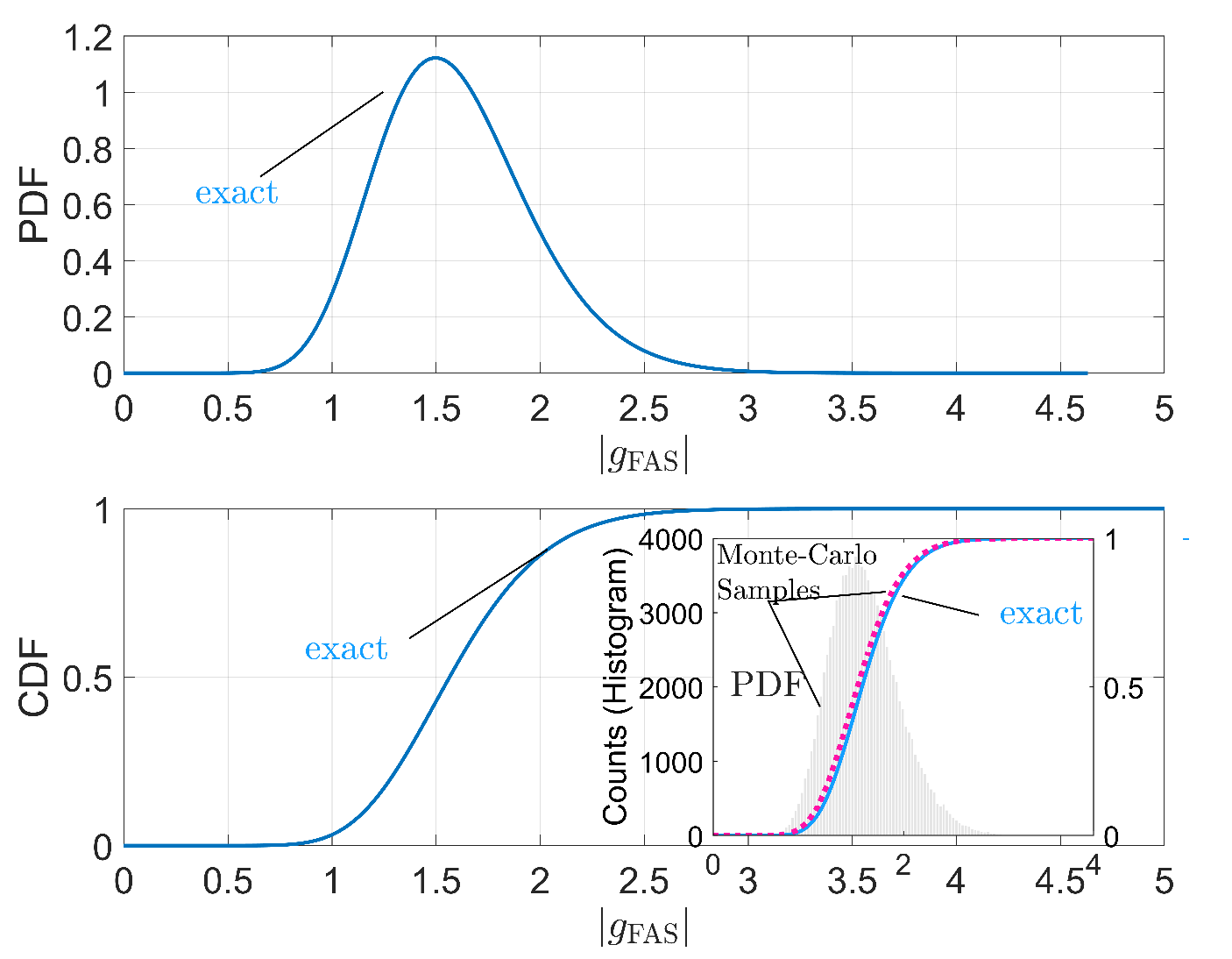}
	\caption{Illustration of the simplified CDF in \eqref{joint CDF} and PDF in \eqref{eq:16} of $|g_{\mathrm{FAS}}|$ with $10^{5}$ Monte-Carlo samples, $\sigma=1$, number of available ports $N=10$ and antenna length constant $W=0.5$.}
	\label{fig:PDF_CDF_Approximation}
\end{figure}
\subsection{BLER Upper-Bound for FBL-FAS}\label{Main Results-B}
In this part, we derive the BLER upper-bounds of FBL-FAS under the case of \textit{known} channel conditions in \textbf{Theorem 1} and the case with \textit{no prior of CSI} in \textbf{Theorem 3}. For ease of description, we rewrite \eqref{eq:5} into a more compact form:
\begin{equation}\label{eq:11}
	\boldsymbol{y}=\sum_{u=1}^{U}g_{u,k}\boldsymbol{x}_u+\boldsymbol{\eta}=\boldsymbol{X}\boldsymbol{g}+\boldsymbol{\eta},
\end{equation}
with codeword (column of $\boldsymbol{X}$) $\boldsymbol{x}_u\in \mathcal{U}$ and we assume the fluid antenna can always switch to the ports with maximum amplitude. Let $g_{\mathrm{FAS}}$ be the random variable denoting the channel response afer port selection, which is defined by:
\begin{equation}\label{eq:12}
	|g_{\mathrm{FAS}}| = \max\left\{|g_{u,1}|,|g_{u,2}|,\ldots,|g_{u,N}|\right\}.
\end{equation}
The correlation among ports has been reported in \eqref{eq:4}, \eqref{eq:modified} and \eqref{eq:full}. One thing worth noting is that though each realization for each users' port selection may be different, the amplitude's statistical behaviors of all selected ports are identical to \eqref{eq:12}.

We first define the error event set ${\mathcal{U}}'$ denoting there are ${U}'$ incorrect detection in each ${\mathcal{U}}'$, i.e., $|{\mathcal{U}}'|={U}'$. Considering all possible combinations of ${\mathcal{U}}'$, the universal set of error event is denoted by $\mathcal{W}=\{{\mathcal{U}}'_i,i=1,2,\ldots\}$. The event $\mathcal{E}_{{U}'}$ denoting ${U}'$ erroneous detections takes place when:
\begin{subequations}\label{eq:13}
	\begin{align}
	\mathcal{E}_{{\mathcal{U}}'}&=\begin{Bmatrix}
	\|\boldsymbol{y}-\underbrace{{\boldsymbol{X}}'{\boldsymbol{g}}'}_{
		\text{${U}'$ error}}-\underbrace{\boldsymbol{X}_{\mathcal{U}\setminus{\mathcal{U}}'}\boldsymbol{g}_{\mathcal{U}\setminus{\mathcal{U}}'}}_{\text{$U-{U}'$ correctly detected}}\|^2_2 \\
	< \|\boldsymbol{y}-\boldsymbol{X}\boldsymbol{g}\|^2_2		
	\end{Bmatrix}, \label{eq:13a}\\
	&=\left\{\|\underbrace{\left(\boldsymbol{X}_{{U}'}-{\boldsymbol{X}}'\right){\boldsymbol{g}}'}_{\boldsymbol{\eta}'}+\boldsymbol{\eta}\|^2_2 < \|\boldsymbol{\eta}\|^2_2 \right\}, \label{eq:13b}
		\end{align}
\end{subequations}
where in \eqref{eq:13a} ${\boldsymbol{X}}'$ are signals in error set ${\mathcal{U}}'$ with corresponding channel coefficients ${\boldsymbol{g}}'$, $\mathcal{U}\setminus{\mathcal{U}}'$ denotes signal set excluding incorrectly detected with cardinality $|\mathcal{U}\setminus{\mathcal{U}}'|=U-{U}'$ and in \eqref{eq:13b} $\boldsymbol{X}_{{U}'}$ is the correct signals corresponding to those incorrectly detected. We denote $\boldsymbol{\eta}'=\left(\boldsymbol{X}_{{U}'}-{\boldsymbol{X}}'\right){\boldsymbol{g}}'$ in the sequel. Meanwhile, there are $|\mathcal{W}|=\binom{U}{{U}'}^2 $ possible combinations for ${\mathcal{U}'}$ \cite[Eq.~13]{chernoff1}, \cite[Eq.~12]{chernoff2}. Thereby, the BLER for FBL-FAS can be calculated as:
\begin{equation}\label{eq:14}
	\mathrm{BLER}=P\left(\bigcup_{{\mathcal{U}}'\in \left\{{\mathcal{U}}'\right\}}\mathcal{E}_{{\mathcal{U}}'}\right)\le |\mathcal{W}|P\left(\mathcal{E}_{{\mathcal{U}}'}\right),
\end{equation}
where the inequity holds since $P\left(A_1+A_2\ldots\right)\le \sum_{i}P\left(A_i\right)$.
\par\indent{\bf\em Theorem 1:} The BLER conditioned on $|g_{\mathrm{FAS}}|$ is upper-bounded by 
\begin{equation}\label{eq:15}
	P\left(\mathcal{E}_{{\mathcal{U}}'}\Big\vert |g_{\mathrm{FAS}}|\right)\le \sum_{{U}'=0}^{U}\frac{{U}'}{U}e^{{L}'-M\log\left(1+\frac{0.25M\sigma_{\eta'}^2}{\sigma_{\eta}^2}\right)},
\end{equation}
where ${L}'=\log\binom{U}{{U}'}^2=\sum_{i=1}^{{U}'-1}2\log\frac{U-i}{{U'-i}}$ and $\sigma^2_{\eta'}=2{U}'\sigma_c^2|g_{\mathrm{FAS}}|^2$.
\par\indent\indent{\bf\em Proof:} See Appendix~\ref{appen-d}. \hfill$\blacksquare$

\par \indent \textit{Note that, \textbf{Theorem 1} is only valid when the channel gain $|g_{\mathrm{FAS}}|$ is known, in which case the conditional error probability is calculated using \eqref{eq:15}.} 

In the following, we investigate the BLER assuming no prior of CSI, the channel gain is unknown at the receiver and no longer deterministic. In this case, $|g_{\mathrm{FAS}}|$ is treated as a random variable, and the corresponding statistical BLER for FBL-FAS is derived. The simple reference correlation model is utilized as a example.
\par\indent{\bf\em Theorem 2:} The PDF of random variable $|g_{\mathrm{FAS}}| = \max\left\{|g_{u,1}|,|g_{u,2}|,\ldots,|g_{u,N}|\right\}$ based on simple reference channel model in Sec.~\ref{sec.simple_reference_model} is given in \eqref{eq:16}:
\begin{equation}\label{eq:16}
	\footnotesize
	\begin{aligned}
		&f_{|g_{\mathrm{FAS}}|}(r) =\\
		&\frac{2r}{\sigma^2} e^{-\frac{r^2}{\sigma^2}} \prod_{k=2}^N \left[ 1 - Q_1\left( \sqrt{\frac{2 \mu_k^2}{1 - \mu_k^2}} \frac{r}{\sigma}, \sqrt{\frac{2}{1 - \mu_k^2}} \frac{r}{\sigma} \right) \right]+\\
		&\sum_{i=2}^N \int_0^{\frac{r^2}{\sigma^2}} e^{-t} \left[ \prod_{k=2, k \neq i}^N \left( 1 - Q_1\left( \sqrt{\frac{2 \mu_k^2}{1 - \mu_k^2}} \sqrt{t}, \sqrt{\frac{2}{\sigma^2 (1 - \mu_k^2)}} r \right) \right) \right] \\
		&\times
		\frac{2r}{1 - \mu_i^2} \frac{1}{\sigma^2} \left( e^{ -\frac{\frac{2 \mu_i^2 t}{1 - \mu_i^2} + \frac{2 r^2}{\sigma^2 (1 - \mu_i^2)}}{2} }\right) I_0\left( \sqrt{\frac{2 \mu_i^2 t}{1 - \mu_i^2}}\sqrt{\frac{2}{1 - \mu_i^2}} \frac{r}{\sigma} \right)\mathrm{d}t.
	\end{aligned}
\end{equation}
\par\indent\indent{\bf\em Proof:} See Appendix~\ref{appen-e}. \hfill$\blacksquare$

In Fig.~\ref{fig:PDF_CDF_Approximation}, we illustrate and compare the exact CDF and PDF with Monte-Carlo samples under setups $\sigma=1$, $N=10$ and $W=0.5$.
\par\indent{\bf\em Theorem 3:} With unknown channel coefficients, the BLER for FBL-FAS can be upper-bounded by:
\begin{equation}\label{eq:17}
		\begin{aligned}
			\mathrm{BLER}&=\int f_{|g_{\mathrm{FAS}}|}(r) P\left(\mathcal{E}_{{\mathcal{U}}'}\Big\vert |g_{\mathrm{FAS}}|\right) \mathrm{d}r\\ 
			&\le\sum_{{U}'=0}^{U}\int_{r=0}^{+\infty}f_{|g_{\mathrm{FAS}}|}(r)\times \frac{{U}'}{U}e^{{L}'-M\log\left(1+\frac{0.25M\sigma_{\eta'}^2}{\sigma_{\eta}^2}\right)} \mathrm{d}r,\\
			&\sigma^2_{\eta'}=2{U}'\sigma_c^2r^2,
			{L}'=\sum_{i=1}^{{U}'-1}2\log\frac{U-i}{{U'-i}}.
		\end{aligned}
\end{equation}
\par\indent\indent{\bf\em Proof:} {\bf Theorem~1} has concluded the BLER conditioned on $|g_{\mathrm{FAS}}|$. And if the PDF $f_{|g_{\mathrm{FAS}}|}(r)$ discussed in {\bf Theorem~2} is further considered, the statistical BLER for FBL-FAS can be derived by integrating $f_{|g_{\mathrm{FAS}}|}(r)$ within channel response domain outputs, which completes the proof. \hfill$\blacksquare$
\subsection{Benchmark: Conventional $L$-Antenna System}\label{sec:benchmark_metric}
	For comparison, we next present the expressions of BLER and SINR for a conventional $L$-antenna system. According to \cite[Eq.~23]{FBL-FAS}, if the channel coefficients are perfectly known, the maximum SINR after maximal ratio combining (MRC) for an $L$-antenna system is:
	\begin{equation}\label{eq:max_sinr}
		\widehat{\gamma}_{L}=\frac{4L^3\sigma^2}{\pi\left(U-1\right)\sigma^2+4L^2\sigma_\eta^2},
	\end{equation} 
	with which the corresponding BLER  can be computed by \cite{intro_FBL1} as:
	\begin{equation}\label{eq:BLER-conventional}
		\epsilon_L=Q\left(\frac{C-R_c}{\sqrt{V_{dis}/M}}\right),
	\end{equation}
	where $C=\frac{1}{2}\log_2\left(1+\widehat{\gamma}_{L}\right)$ is the averaged channel capacity, $V_{dis}=\frac{\widehat{\gamma}_L}{2}\frac{\widehat{\gamma}_L+2}{(\widehat{\gamma}_L+1)^2}\log_2^2(e)$ denotes the channel dispersion and $R_c=\frac{\log_2U}{M}$ is the code rate. Closed-from BLER for conventional systems in \cite{intro_FBL1} offers accurate performance prediction when blocklength is relatively large in the order of $10^2$.
\begin{figure}[t!]
	\centering
	\includegraphics[width=3in]{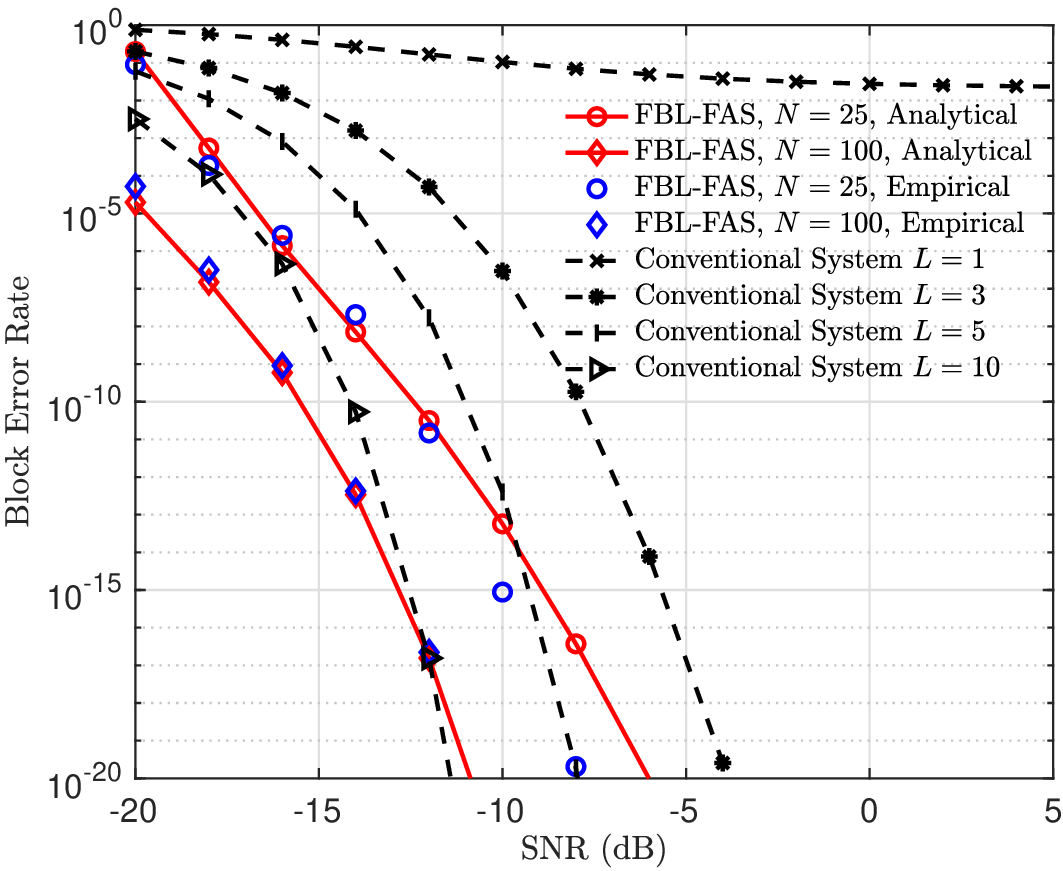}
	\caption{BLER performance versus SNR (dB). Configurations: $U=20$, $M=400$, $W=2$, $N\in\{5,25\}$, $L\in\{1,3,5,10\}$.}
	\label{fig:BLER_SNR}
\end{figure}
\begin{figure}[t!]
	\centering
	\includegraphics[width=3in]{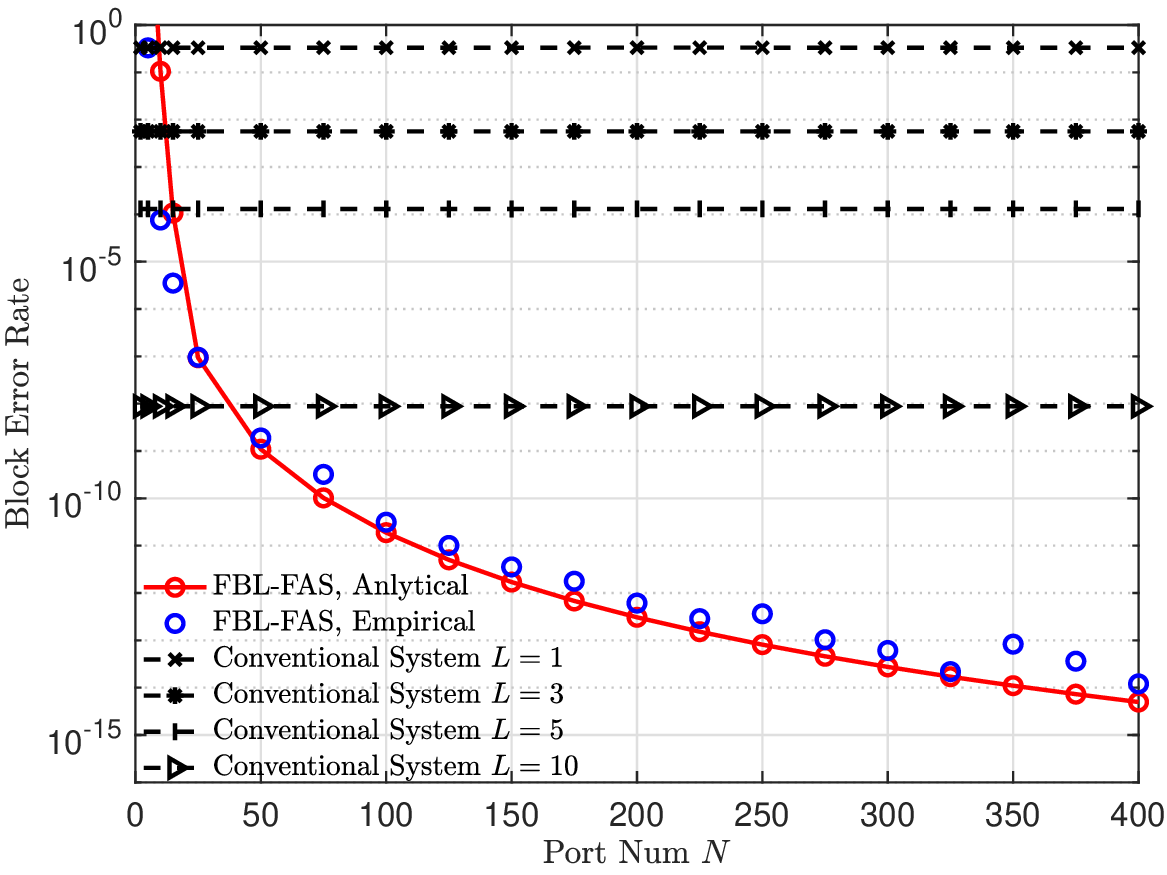}
	\caption{BLER performance versus port number $N$. Configurations: $U=20$, $M=400$, $W=2$, $\mathrm{SNR}=-15$ dB, $L\in\{1,3,5,10\}$.}
	\label{fig:BLER_N}
\end{figure}
\begin{figure}[t!]
	\centering
	\includegraphics[width=3in]{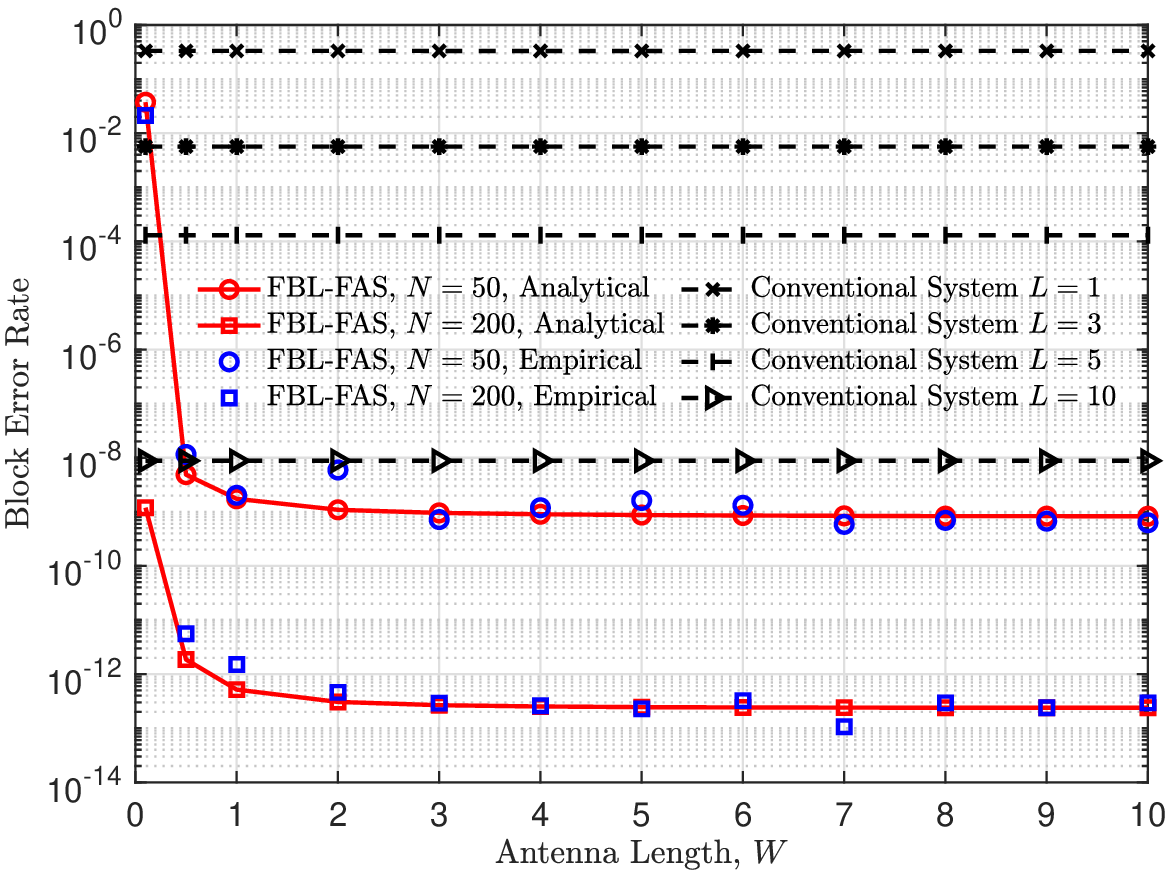}
	\caption{BLER performance versus number of codeword $U$. Configurations:  $M=400$, $\mathrm{SNR}=-15$ dB, $N\in\{50,200\}$, $L\in\{1,3,5,10\}$.}
	\label{fig:BLER_W}
\end{figure}
\section{Numerical Results}\label{sec: numerical results}
In this section, we present and compare the BLER performance of different systems, namely, the FBL-FAS and the conventional $L$-antenna system under finite-blocklength region. Throughout the simulations, the FAS configuration employs only a single RF chain, whereas the $L$-antenna system is assumed to use $L$ RF chains, i.e., $L$ independent antennas for signal reception in conventional system with fixed antenna.
\subsection{BLER under Simple Reference Correlation Model}
Fig.~\ref{fig:BLER_SNR} illustrates the BLER across various SNR (dB) ranges with $U=20$, $M=400$, $W=2$, $N\in\{25,100\}$, and $L\in\{1,3,5,10\}$. An interesting observation is that the single-antenna conventional system suffers from a performance error floor, which originates from multi-user interference and is also predictable according to \eqref{eq:max_sinr}. FAS demonstrates substantial performance gains, e.g., for $N=25$, the single-RF-chain FAS yields a significantly lower BLER compared with the 5-antenna conventional system, and for $N=100$, it achieves an almost $10^{5}$-fold BLER reduction relative to $N=25$.

Fig.~\ref{fig:BLER_N} depicts the BLER performance for different numbers of ports $N$ under $U=20$, $M=400$, $W=2$, $\mathrm{SNR}=-15$~dB, and $L\in\{1,3,5,10\}$. As the number of available ports increases, the potential spatial diversities become substantial. After $N=50$, the FBL-FAS with a single activated port achieves a lower BLER than the conventional antenna system with $L=10$ antennas. Meanwhile, the empirical results exhibit close alignment with the theoretical prediction.

Moreover, Fig.~\ref{fig:BLER_W} illustrates the BLER performance under different array length $W$, with parameters $M=20$, $W=2$, $\mathrm{SNR}=-15$~dB, $N\in\{50,200\}$, and $L\in\{1,3,5,10\}$. With $50$ available ports, the FBL-FAS achieves performance comparable to the 10-antenna conventional system under a $W=0.5$-length array. With increased $N$, the BLER performance is further enhanced due to the improved spatial diversities.
\begin{figure}[t!]
	\centering
	\includegraphics[width=3in]{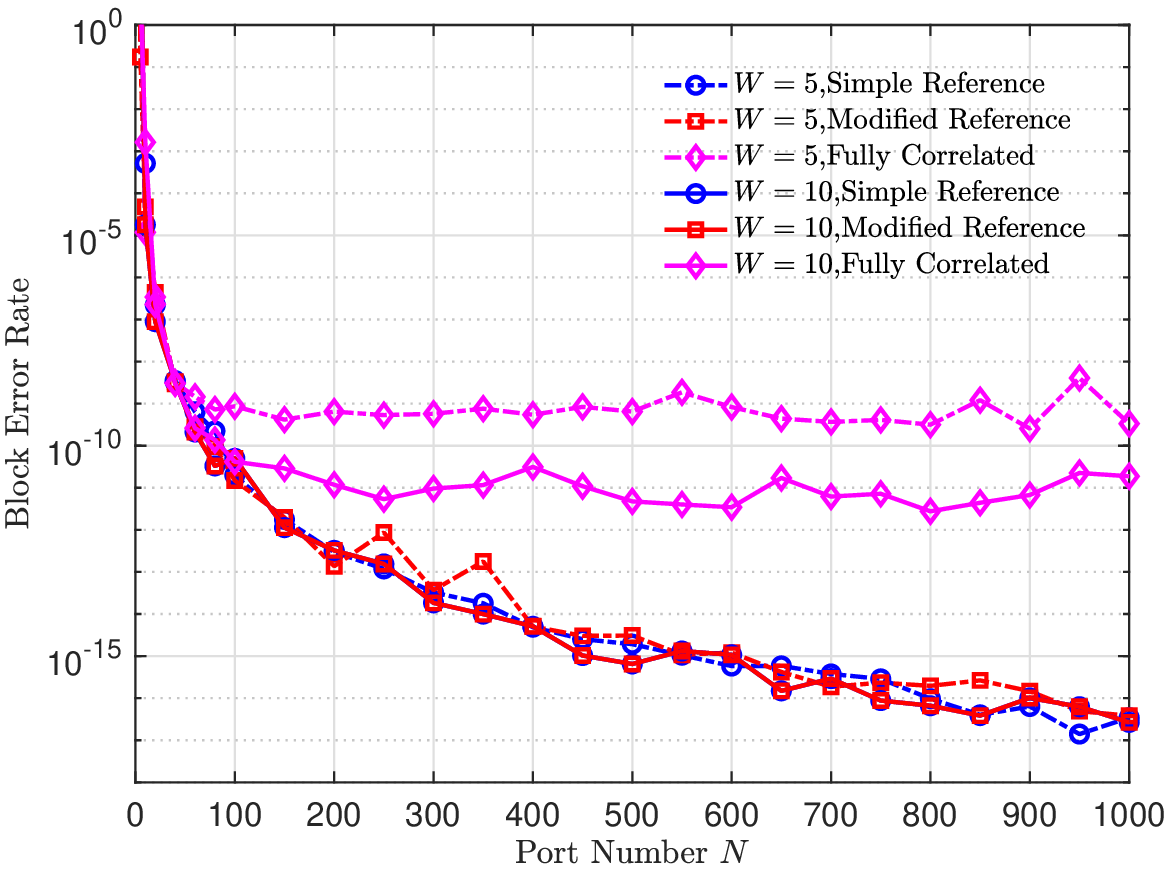}
	\caption{BLER performance under different correlation models versus port number $N$. Configurations: $M=400$, $U=20$, $W\in\{5,10\}$, $\mathrm{SNR}=-15$ dB.}
	\label{fig:BLER_models}
\end{figure}
\subsection{BLER Comparison under Different Correlation Models}
In this section, the BLER performance under different correlation models is compared to demonstrate the universality of the proposed bound and to highlight the distinctions among various correlation models. As shown in Fig.~\ref{fig:BLER_models}, the BLER performance is evaluated under the settings $M=400$, $U=20$, $W\in\{5,10\}$, and $\mathrm{SNR}=-15$~dB.

Several observations can be made. The fully correlated model, capturing all inter-port correlations, exhibits the strongest correlation and thus the poorest BLER performance. The simple and modified reference models behave similarly due to their limited modeling precision. Overall, when the number of available ports is small (e.g., before $N=100$ in Fig.~\ref{fig:BLER_models}), all models yield nearly identical BLER since the correlation structure has limited impact. As $N$ grows and ports become densely distributed, model differences become pronounced: reference-based models tend to produce overly optimistic predictions, while the fully correlated model consistently provides the most conservative BLER. Despite this, all models exhibit the same overall performance trend, confirming that the inherent advantage of FAS is preserved regardless of the correlation model used. {\em Nonetheless, for more accurate performance evaluation at large $N$, the fully correlated or block-correlation models~\cite{block_FBL_FAS,block1} are recommended.}
\section{Conclusion}\label{conclusion}
This work introduces how the finite blocklength effect can be incorporated into FAS. More importantly, a universal BLER performance bound is presented, which can be effectively applied to various FAS designs with different correlation models. The proposed BLER bound is applicable in both model-aware scenarios, such as reference-based correlation models and their variance, as well as in model-free scenarios, such as the fully correlated channel model. Furthermore, in model-aware cases, the analytical results closely match the empirical data, highlighting the versatility of the proposed bound. Overall, the proposed BLER bound provides a solid theoretical benchmark for practical FAS designs considering finite blocklength.
\section{Appendix}\label{appendx}
\subsection{Probability of BLER Conditioned on $|g_{\mathrm{FAS}}|$}\label{appen-d}
In this part, we first confirm the elements' distribution of $\boldsymbol{\eta}'=\left(\boldsymbol{X}_{{U}'}-{\boldsymbol{X}}'\right){\boldsymbol{g}}'$ in \eqref{eq:13b}. Subsequently, we derive the BLER conditioned on $|g_{\mathrm{FAS}}|$ and with the derived PDF of $|g_{\mathrm{FAS}}|$, the BLER upper-bound is derived.

For the $m$-th element $\eta_m',~m=1,2,\ldots,M$ in $\boldsymbol{\eta}'=\left(\boldsymbol{X}_{{U}'}-{\boldsymbol{X}}'\right){\boldsymbol{g}}'$, $\eta_m'=\sum_{u=1}^{{U}'}\left(c_{u_i,m}-c_{u_j,m}\right)g_{\mathrm{FAS}}$ where $c_{u_i,m}\neq c_{u_j,m}$ are elements from two arbitrary different codewords. Considering $c_{u_i,m},c_{u_j,m}\sim \mathcal{CN}\left(0,\sigma_c^2\right)$ are independent, the distribution conditioned on $|g_{\mathrm{FAS}}|$ is $\eta_m'\sim \mathcal{CN}\left(0,\underbrace{2{U}'\sigma_c^2|g_{\mathrm{FAS}}|^2}_{\sigma^2_{\eta'}} \Big\vert |g_{\mathrm{FAS}}|\right)$ and we use $\sigma^2_{\eta'}=2{U}'\sigma_c^2|g_{\mathrm{FAS}}|^2$ in the sequel.

Here, we explain how to derive the BLER upper-bound. Conditioned on $|g_{\mathrm{FAS}}|$:
	\begin{align}\label{BLER-conditioned}
			P\left(\mathcal{E}_{{\mathcal{U}}'}\Big\vert |g_{\mathrm{FAS}}|\right) =P\left(\|\boldsymbol{\eta}'+\boldsymbol{\eta}\|^2_2 < \|\boldsymbol{\eta}\|^2_2 \Big\vert |g_{\mathrm{FAS}}| \right).
	\end{align}
To proceed with more conciseness, the condition notation $\left(\left(\cdot\right) \Big\vert |g_{\mathrm{FAS}}|\right)$ is omitted unless particularly stated. By utilizing Chernoff Inequity $P\left(x>0\right)\le \mathrm{E}\left[e^{\lambda_1x}\right]$ \cite{chernoff2,chernoff3,chernoff1}, \eqref{BLER-conditioned} is converted into:
\begin{subequations}\label{chernoff}
	\begin{align}
		&P\left(\|\boldsymbol{\eta}'+\boldsymbol{\eta}\|^2_2 <\|\boldsymbol{\eta}\|^2_2 \right)\le \mathrm{E}\left[e^{-\lambda_1\|\boldsymbol{\eta}'+\boldsymbol{\eta}\|_2^2+\lambda_1\|\boldsymbol{\eta}\|_2^2}\right] \label{chernoff-a}\\
		&=\mathrm{E}\left[\frac{e^{\lambda_1\|\boldsymbol{\eta}\|^2_2}e^{\frac{-\lambda_1\|\boldsymbol{\eta}\|^2_2}{1+\lambda_1\sigma^2_{\eta'}}}}{\left(1+\lambda_1\sigma^2_{\eta'}\right)^M}\right]=
		\mathrm{E}\left[\frac{e^{\left(\frac{\lambda^2_1\sigma^2_{\eta'}}{(1+\lambda_1\sigma_{\eta'}^2)}\right)\|\boldsymbol{\eta}\|_2^2}}{\left(1+\lambda_1\sigma^2_{\eta'}\right)^M}\right]\label{chernoff-b}\\
		&
		=\frac{1}{\left(1+\lambda_1\sigma^2_{\eta'}\right)^M}\frac{1}{\left(1-\frac{\lambda^2_1\sigma^2_{\eta'}\sigma_{\eta}^2/M}{(1+\lambda_1\sigma_{\eta'}^2)}\right)^M}\label{chernoff-c}\\
		&=\frac{1}{\left(1+\lambda_1\sigma_{\eta'}^2-\lambda_1^2\sigma_{\eta'}^2\sigma_{\eta}^2/M\right)^{M}}\xrightarrow[\text{denominator maximization}]{\lambda_1=0.5M/\sigma_{\eta}^2}\label{chernoff-d}\\
		&=e^{-M\log\left(1+\frac{0.25M\sigma_{\eta'}^2}{\sigma_{\eta}^2}\right)},\label{chernoff-e}
	\end{align}
\end{subequations}
where from \eqref{chernoff-b} to \eqref{chernoff-c}, identity of $\mathrm{E}\left[e^{x\|\boldsymbol{a}+\boldsymbol{b}\|_2^2}\right]=\frac{e^{\frac{x\|\boldsymbol{b}\|_2^2}{\left(1-x\sigma_a^2\right)}}}{\left(1-x\sigma_a^2\right)^M}$ is adopted twice. Thereby, considering \eqref{eq:14}, the BLER conditioned on $|g_{\mathrm{FAS}}|$ is upper-bounded by $P\left(\mathcal{E}_{{\mathcal{U}}'}\Big\vert |g_{\mathrm{FAS}}|\right)\le \sum_{{U}'=0}^{U}\frac{{U}'}{U}e^{{L}'-M\log\left(1+\frac{0.25M\sigma_{\eta'}^2}{\sigma_{\eta}^2}\right)}$, where ${L}'=\log\binom{U}{{U}'}^2=\sum_{i=1}^{{U}'-1}2\log\frac{U-i}{{U'-i}},~
\sigma^2_{\eta'}=2{U}'\sigma_c^2|g_{\mathrm{FAS}}|^2$ and ratio $\frac{{U}'}{U}$ denotes the BLER when ${U}'$ error exists, which completes the proof for {\bf Theorem 1} in \eqref{eq:15}.

\subsection{Derivations of Random Variable $|g_{\mathrm{FAS}}|$'s PDF $f_{|g_{\mathrm{FAS}}|}\left(r\right)$}\label{appen-e}
With conditional probability of BLER in proof~\ref{appen-d}, we can finalize the upper-bounding on BLER of FBL-FAS with PDF of $|g_{\mathrm{FAS}}|$. We denote the PDF of $|g_{\mathrm{FAS}}|$ by $f_{|g_{\mathrm{FAS}}|}\left(r\right)$. Some useful remarks are listed:
\par \indent{\bf \em Remark E1:} \textit{According to \cite{fas-twc-21}, the joint PDF of $|g_1|,|g_2|,|g_3|,\ldots,|g_N|$ is:}
\begin{equation}
	\begin{aligned}
		&p_{|g_1|,\ldots,|g_N|}\left(r_1,\ldots,r_N\right), r_1,\ldots,r_N\ge 0\\
		&=\prod_{k=1}^N \frac{2 r_k}{\sigma^2 (1 - \mu_k^2)} e^{ \left( -\frac{r_k^2 + \mu_k^2 r_1^2}{\sigma^2 (1 - \mu_k^2)} \right)} I_0\left( \frac{2 \mu_k r_1 r_k}{\sigma^2 (1 - \mu_k^2)} \right),
	\end{aligned}
\end{equation}
\textit{and the joint CDF of $|g_1|,|g_2|,|g_3|,\ldots,|g_N|$ is:}
\begin{equation}\label{eq:joint_cdf_appedx}
	\begin{aligned}
&C_{|g_1|, \dots, |g_N|}(r_1, \dots, r_N)\\
&= \int_0^{\frac{r_1^2}{\sigma^2}} e^{-t} \times\\
& \quad \prod_{k=2}^N \left[ 1 - Q_1\left( \sqrt{\frac{2 \mu_k^2}{1 - \mu_k^2}} \sqrt{t}, \sqrt{\frac{2}{\sigma^2 (1 - \mu_k^2)}} r_k \right) \right] dt,
	\end{aligned}
\end{equation}
\textit{where $I_0\left(\cdot\right)$ is the zero-order modified Bessel function of the first kind and $Q_1\left(a,b\right)$ is the first-order Marcum $Q$-function.}

With {\bf Remark E1}, it is feasible to see that the CDF of random variable $|g_{\mathrm{FAS}}|$ is:
\begin{equation}\label{joint CDF}
	\footnotesize
	\begin{aligned}
		&C_{|g_{\mathrm{FAS}}|}(r) = P(|g_1| \leq r, \dots, |g_N| \leq r)\\
		& = C_{|g_1|, \dots, |g_N|}(r_1=r, \dots, r_N=r)\\
		& = \int_0^{\frac{r^2}{\sigma^2}} e^{-t} \prod_{k=2}^N \left[ 1 - Q_1\left( \sqrt{\frac{2 \mu_k^2}{1 - \mu_k^2}} \sqrt{t}, \sqrt{\frac{2}{\sigma^2 (1 - \mu_k^2)}} r \right) \right] \mathrm{d}t.
	\end{aligned}
\end{equation}
Thereby, the $|g_{\mathrm{FAS}}|$'s PDF is:
\begin{equation}
	f_{|g_{\mathrm{FAS}}|}\left(r\right)=\frac{\mathrm{d}}{\mathrm{d}r} C_{|g_{\mathrm{FAS}}|}(r).
\end{equation}
Utilizing Leibniz rule of $\frac{\mathrm{d}}{\mathrm{d}r} \int_{a(r)}^{b(r)} g(t, r) \, \mathrm{d}t = g(b(r), r) \frac{\mathrm{d}b}{\mathrm{d}r} - g(a(r), r) \frac{\mathrm{d}a}{\mathrm{d}r} + \int_{a(r)}^{b(r)} \frac{\partial g(t, r)}{\partial r} \, \mathrm{d}t$, and substituting $u = \frac{r^2}{\sigma^2},\frac{du}{dr} = \frac{2r}{\sigma^2}$, one will have PDF in \eqref{d_FAS_PDF}:
	\begin{equation}\label{d_FAS_PDF}
		\footnotesize
		\begin{aligned}
			&f_{|g_{\mathrm{FAS}}|}(r)=\\
			 &\frac{2r}{\sigma^2} e^{-\frac{r^2}{\sigma^2}} \prod_{k=2}^N \left[ 1 - Q_1\left( \sqrt{\frac{2 \mu_k^2}{1 - \mu_k^2}} \frac{r}{\sigma}, \sqrt{\frac{2}{1 - \mu_k^2}} \frac{r}{\sigma} \right) \right] +\\
			&\sum_{i=2}^N \int_0^{\frac{r^2}{\sigma^2}} e^{-t} \left( \prod_{k=2, k \neq i}^N \left[ 1 - Q_1\left( \sqrt{\frac{2 \mu_k^2}{1 - \mu_k^2}} \sqrt{t}, \sqrt{\frac{2}{1 - \mu_k^2}} \frac{r}{\sigma} \right) \right] \right)\\ \times
			& \underbrace{\frac{\partial}{\partial r} \left[ 1 - Q_1\left( \sqrt{\frac{2 \mu_i^2}{1 - \mu_i^2}} \sqrt{t}, \sqrt{\frac{2}{1 - \mu_i^2}} \frac{r}{\sigma} \right) \right] \mathrm{d}t}_{\eqref{d_component}},
		\end{aligned}
	\end{equation}
where the differentiation term can be calculated in \eqref{d_component}:
	\begin{equation}\label{d_component}
			\footnotesize
		\begin{aligned}
			&\frac{\partial}{\partial r} \left[ 1 - Q_1\left( \sqrt{\frac{2 \mu_i^2}{1 - \mu_i^2}} \sqrt{t}, \sqrt{\frac{2}{1 - \mu_i^2}} \frac{r}{\sigma} \right) \right]= \\
			&\frac{2r}{1 - \mu_i^2} \frac{1}{\sigma^2} \left(e^{ -\frac{\frac{2 \mu_i^2 t}{1 - \mu_i^2} + \frac{2 r^2}{\sigma^2 (1 - \mu_i^2)}}{2} }\right)\times I_0\left( \sqrt{\frac{2 \mu_i^2 t}{1 - \mu_i^2}} \cdot \sqrt{\frac{2}{1 - \mu_i^2}} \frac{r}{\sigma} \right).
		\end{aligned}
	\end{equation}
Further, considering the following entities
\begin{equation}
	\begin{aligned}
		\frac{\partial}{\partial b} Q_1(a, b) &= -b e^{\left( -\frac{a^2 + b^2}{2} \right)} I_0(ab),\\
		\frac{\partial}{\partial r} Q_1(a \sqrt{t}, b r) &= b^2 e^{\left( -\frac{a^2 t + b^2 r^2}{2} \right)} I_0(a \sqrt{t} \cdot b r),
	\end{aligned}
\end{equation} 
and substituting $ a = \sqrt{\frac{2 \mu_i^2}{1 - \mu_i^2}} \sqrt{t} $, $ b = \sqrt{\frac{2}{1 - \mu_i^2}} \frac{1}{\sigma}$, \eqref{d_component} is obtained. Subsequently, substituting \eqref{d_component} into \eqref{d_FAS_PDF} concludes the PDF in \eqref{eq:16}, which completes the proof for {\bf Theorem 2} in \eqref{eq:16}.

\balance

%
%
%


%
%
%
%
%

\end{document}